\documentclass{aastex631}
\accepted{October 3, 2023}
\submitjournal{RNAAS}
\usepackage{upquote}

\shorttitle{DRAGONS v3.1}
\shortauthors{DRAGONS Team}

\begin{document}

\title{DRAGONS - A Quick Overview}

\author[0000-0002-6633-7891]{K. Labrie}
\affil{Gemini Observatory/NSF’s NOIRLab, 670 N. A’ohoku Place, Hilo, Hawai’i, 96720, USA}
\email{kathleen.labrie@noirlab.edu}

\author[0000-0001-8589-4055]{C. Simpson}
\affil{Gemini Observatory/NSF’s NOIRLab, 670 N. A’ohoku Place, Hilo, Hawai’i, 96720, USA}

\author{R. Cardenes}
\affil{Gemini Observatory/NSF’s NOIRLab, Casilla 603, La Serena, Chile}

\author{J. Turner}
\affil{Gemini Observatory/NSF’s NOIRLab, Casilla 603, La Serena, Chile}

\author[0000-0001-6360-992X]{M. Soraisam}
\affil{Gemini Observatory/NSF’s NOIRLab, 670 N. A’ohoku Place, Hilo, Hawai’i, 96720, USA}

\author[0000-0002-1557-3560]{B. Quint}
\affil{Rubin Observatory Project Office, 950 N. Cherry Ave., Tucson, AZ 85719, USA}

\author{O. Oberdorf}
\affil{Gemini Observatory/NSF’s NOIRLab, 670 N. A’ohoku Place, Hilo, Hawai’i, 96720, USA}

\author[0000-0003-4479-1265]{V. M.\ Placco}
\affiliation{NSF’s NOIRLab, 950 N. Cherry Ave., Tucson, AZ 85719, USA}

\author[0000-0002-7181-3020]{D. Berke}
\affil{Gemini Observatory/NSF’s NOIRLab, 670 N. A’ohoku Place, Hilo, Hawai’i, 96720, USA}

\author{O. Smirnova}
\affil{Gemini Observatory/NSF’s NOIRLab, Casilla 603, La Serena, Chile}

\author[0000-0002-3657-4191]{S. Conseil}
\affil{Aix Marseille Univ, CNRS, CNES, LAM, Marseille, France}

\author[0000-0002-9123-0068]{W. D. Vacca}
\affil{Gemini Observatory/NSF’s NOIRLab, 950 N. Cherry Avenue, Tucson, AZ 85719, USA}

\author[0000-0003-1033-4402]{J. Thomas-Osip}
\affil{Gemini Observatory/NSF’s NOIRLab, Casilla 603, La Serena, Chile}

\begin{abstract}
DRAGONS (Data Reduction for Astronomy from Gemini Observatory North and South) is a platform for the reduction and processing of astronomical data. The Python-based, open-source package includes infrastructure for automation and algorithms for the processing of imaging and spectroscopic data, up to the analysis-ready stage. DRAGONS currently focuses on the reduction of Gemini data, although it allows for support of data from other instruments and telescopes through third-party extensions.  Its latest release (v3.1) enables automated reduction of all currently-active Gemini imaging facility instruments, as well as optical longslit spectroscopic data, acquired with GMOS.
\end{abstract}


\section{DRAGONS} \label{dragons}

DRAGONS is an automated, yet customizable pipeline for the processing of raw astronomical data.  As of v3.1, the software supports science-quality data reduction from all current facility imagers at Gemini, as well as GMOS (optical) longslit spectroscopy data.  The infrastructure and algorithms follow guiding principles that maximize code reuse, minimize long-term maintenance and cost, and streamline development for new instruments.

The automation and pipeline decision-making is critically reliant on the metadata, in particular the header keywords.  The AstroData software (Section \ref{astrodata}), currently distributed with DRAGONS, provides the DRAGONS infrastructure and algorithms, and the calibration association manager with a uniform interface to this information.   

The automated calibration association is supported via the Gemini Observatory Archive (GOA) \citep{GOA:2016} and a calibration manager tool derived from the GOA.  Queries can be made to a light-weight local database or to the online archive.  The best processed calibrations for an observation are automatically retrieved by matching the AstroData tags and descriptors (see below), as prescribed by the calibration association rules defined in the GOA.  

Each individual reduction step is called a `primitive' and a sequence of primitives is a
`recipe'.  While standard reductions are fully automated, the user can modify input parameters to the primitives called by the recipes.  A fork of the LSST project's \verb"pex.config" package \citep{2022:lsstpipe} is used for the input parameter configuration.

The image reductions include instrumental calibrations such as dark, bias, flat, and fringe corrections, as appropriate for the instrument.  For near-infrared data, the sky subtraction pattern (dither-on-target or offset-to-sky) is automatically recognized and the sky frames are assigned to the target frames appropriately, without user input.  The sky subtraction is a two-pass process, where objects are identified and masked in the sky frames, then the frames are aligned, re-sampled, and stacked to generate the final product.

\begin{figure}[ht]
    \centering
    \includegraphics[scale=0.4]{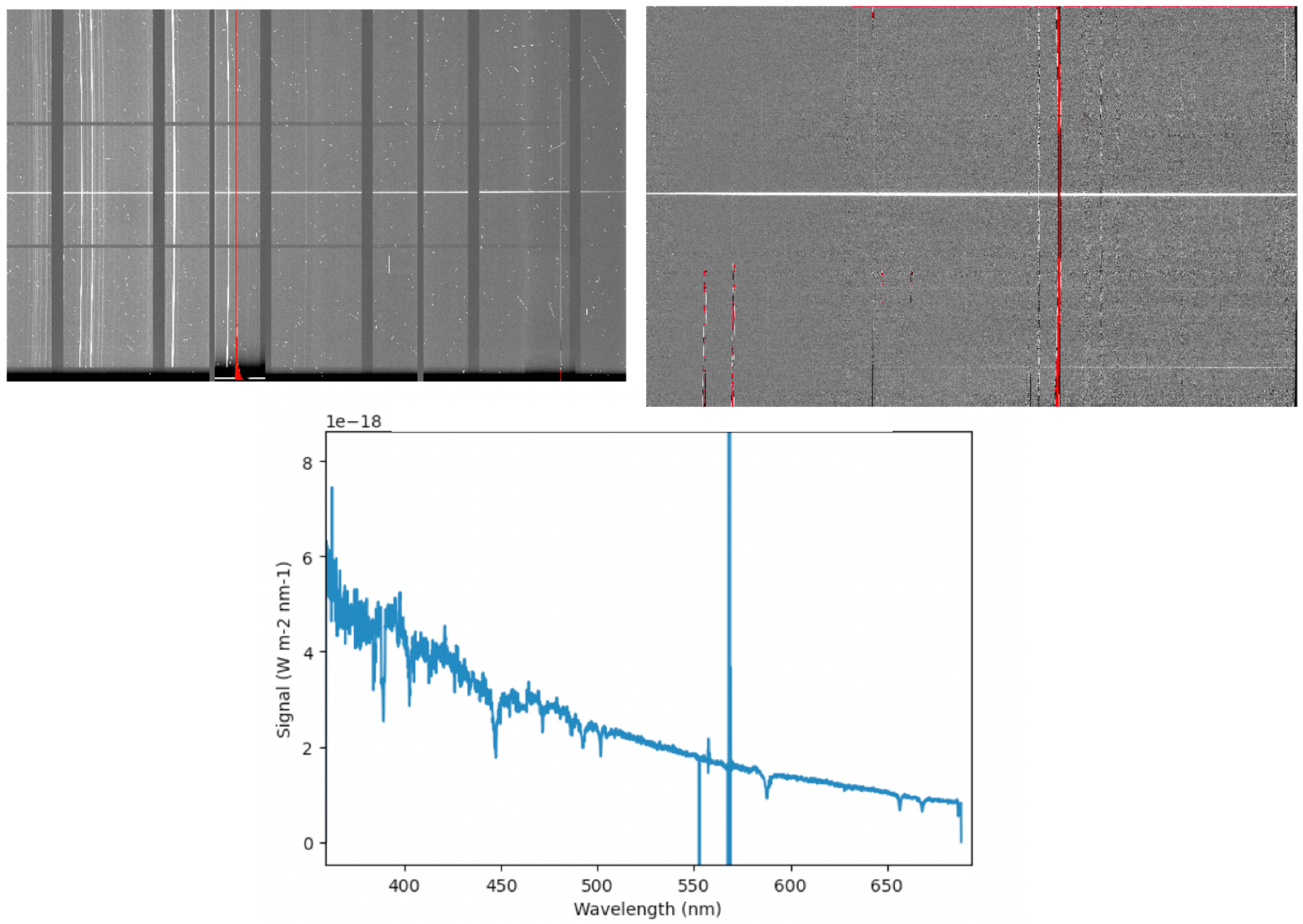}
    \caption{Screenshots of a raw GMOS longslit spectroscopic observation (top-left), a 2-D spectrum from a corrected and stacked 4-frame sequence with spatial and wavelength offsets (top-right), and the final extracted spectrum of the central source (bottom). The source is a DB white dwarf candidate.}
    \label{fig:gmosls}
\end{figure}

Longslit optical spectroscopy processing includes instrumental corrections, sky line removal, wavelength calibration, distortion correction, cosmic ray flagging, alignment and stacking (either 1-D or 2-D spectra), the extraction of the sources detected in the 2-D spectra, and flux calibration.  An example of the products is shown in Figure \ref{fig:gmosls}.

The data in Figure \ref{fig:gmosls} were produced with the shell commands below.  As can be seen, the process is straightforward and accessible to novice users.  An API is also available, allowing users to design their automation strategy.

\begin{verbatim}
    # Create lists with AstroData tags and descriptors
    dataselect ../rawdata/*.fits --tags BIAS \
        --expr='detector_roi_setting=="Central Spectrum"' -o biasesstd.lis
    dataselect ../rawdata/*.fits --tags BIAS \
        --expr='detector_roi_setting=="Full Frame"' -o biasessci.lis
    dataselect ../rawdata/*.fits --tags FLAT -o flats.lis
    dataselect ../rawdata/*.fits --tags ARC -o arcs.lis
    dataselect ../rawdata/*.fits --tags STANDARD -o std.lis
    dataselect ../rawdata/*.fits --xtags CAL --expr='object=="J2145+0031"' -o sci.lis

    # Configure the calibration manager
    vi ~/.dragons/dragonsrc
        [calibs]
        databases = /Local/path/on/user/computer/calmgr.db get store
    
    # Initialize the calibration manager
    caldb init
    
    # Reduce the calibrations, add them to the calibration manager
    # and finally reduce the science data.  Note the "quicklook" (ql)
    # mode is used here.
    caldb add bpm_20140601_gmos-s_Ham_22_full_12amp.fits
    reduce @biasesstd.lis
    reduce @biasessci.lis
    reduce @flats.lis
    reduce @arcs.lis
    reduce @std.lis
    reduce @sci.lis
\end{verbatim}

A suite of browser-based interactive tools, implemented with \verb"bokeh", is available to allow users additional optimization during key spectroscopic reduction steps.

\section{AstroData} \label{astrodata}
A component critical to the automation of DRAGONS is the \verb"astrodata" Python package; it is currently included in DRAGONS.  The AstroData software provides a uniform internal representation of data from different instruments.  The AstroData \emph{tags} and \emph{descriptors} identify and describe the data, allowing DRAGONS to automatically find the matching data reduction recipe library and a matching set of primitives.  The uniform internal representation of the data, from headers to pixels, is central to the maximum code reuse model adopted in DRAGONS.  The \verb"AstroData" class builds upon the \verb"Astropy" \verb"NDData" class \citep{astropy:2018}, with additional attributes and methods to handle memory-mapping of files on disk, and the ability to represent multiple pixel arrays along with their variance and mask.

\section{Upcoming developments} \label{upcoming}
Support for the Gemini near-infrared spectrographs, GNIRS and FLAMINGOS-2, is under active development, with a focus on the longslit and cross-dispersed modes. Once completed, DRAGONS will have tools to support the four basic types of observation: imaging and longslit spectroscopy, in the optical and the near-infrared.   The data reduction software for SCORPIO, the next Gemini imager and spectrograph, will reuse a large fraction of the core DRAGONS primitives.  The longer-term goal is to provide support for all Gemini instruments and modes, including IFU and MOS modes, building upon the basic imaging and spectroscopic algorithms.

The GHOST high-resolution spectrograph data reduction software, initially delivered as a third-party software package, is currently being integrated into core DRAGONS for a seamless user experience.

DRAGONS constitutes an integrated and automated system for reducing data from Gemini instruments. It is extensible and therefore could be adopted by both current and future telescopes, as a foundation for processing the data obtained with their instrumentation.

\section{Resources} \label{resources}
The software is distributed as a conda package under the name \verb|dragons|.  The code and released packages are also available on GitHub at \verb"https://github.com/GeminiDRSoftware/DRAGONS".  The DRAGONS documentation is found at \verb"https://dragons.readthedocs.io/".

\begin{acknowledgments}
The international Gemini Observatory, a program of NSF’s NOIRLab, is managed by the Association of Universities for Research in Astronomy (AURA) under a cooperative agreement with the National Science Foundation on behalf of the Gemini partnership: the National Science Foundation (United States), the National Research Council (Canada), Agencia Nacional de Investigaci\'on y Desarrollo (Chile), Ministerio de Ciencia, Tecnolog\'ia e Innovaci\'on (Argentina), Minist\'erio da Ci\^encia, Tecnologia, Inova\c c\~oes e Comunica\c c\~oes (Brazil), and Korea Astronomy and Space Science Institute (Republic of Korea).

\end{acknowledgments}
\facilities{Gemini:South, Gemini:Gillett}

\software{DRAGONS DOI:10.5281/zenodo.4025470,
Astropy \citep{astropy:2013, astropy:2018}
}

\bibliographystyle{aasjournal}
\bibliography{dragons3AASRN}

\clearpage

\end{document}